\documentclass[twocolumn,aps,showpacs,prl]{revtex4}
\usepackage{amssymb}

\usepackage{epsfig}
\usepackage[english]{babel}
\usepackage{latexsym}
\usepackage{graphics}
\usepackage{subfigure}
\usepackage{epsfig}
\usepackage{graphicx}
\usepackage{dcolumn}
\usepackage{amsmath}

\newcommand{\be}{\begin{equation}}
\newcommand{\ee}{\end{equation}}
\newcommand{\bey}{\begin{eqnarray}}
\newcommand{\eey}{\end{eqnarray}}
\newcommand{\bw}{\begin{widetext}}
\newcommand{\ew}{\end{widetext}}

\newcommand{\ra}{\rangle}

\begin{document}

\title{Fidelity of a Bose-Einstein Condensate}
\author{Jie Liu$^{1,2}$, Wenge Wang$^{1,3}$, Chuanwei Zhang$^{4,5}$, Qian Niu$^{4}$ and Baowen Li$^{1}$}
\date{today}

\begin{abstract}
We investigate fidelity, the Loschmidt echo, for a Bose-Einstein Condensate.  It is found that the fidelity decays
with time in various ways (exponential, Gaussian, and power-law), depending on the choice of initial coherent
state as well as the parameters that determine properties of the underlying classical dynamics. Moreover, high
fidelity is found for initial states lying in the regular region of a mixed-type phase space. A possible
experimental scheme is suggested.
\end{abstract}

 \affiliation{ 
$^1$Department of Physics, National University of Singapore,
117542, Republic of Singapore\\
$^2$Institute of Applied Physics and Computational Mathematics, P.O.Box
100088, Beijing, P.~R.~China\\
$^3$Department of Physics, Southeast University, Nanjing 210096, P.~R.~China
\\
$^4$ Department of Physics, The University of Texas, Austin, Texas
78712-1081 USA \\
$^5$ Center for Nonlinear Dynamics, The University of Texas,
Austin, Texas 78712-1081 USA}  \maketitle

The investigation of coherent manipulation of quantum state of
matter and light has provided insights in many quantum phenomena
and in quantum information processes \cite{lukin}. The realization
of Bose-Einstein condensation (BEC) in dilute gases has provided a
new tool for such investigations \cite{bec}. Recently, it is found
that the stability of quantum state has played a key role in many
procedures for coherent manipulating and applying
BEC\cite{dyinst,Landinst,modins,smerzianglin,liukick}. In fact,
how  to sustain the coherence among the cooled atoms is very
essential for the possible application of BEC to quantum
information and quantum computation.

 However, an important issue is still missing in the study of
 BEC, namely, the sensitivity of the quantum evolution of a BEC
 with respect to the small perturbation that may naturally arise
 from either the manipulation parameters or the interaction with
 environment. This type of stability of quantum motion
 is characterized by the so-called
 fidelity, or the Loschmidt echo, which is defined as the overlap
 of two states obtained by evolving the same initial state under
 two slightly different (perturbed and unperturbed) Hamiltonians.
 This quantity is of special interest in the fields of quantum
 information\cite{quancomp} and  quantum
 chaos\cite{JP01,CT02,JAB02,PZ02,wclp04}.

 In this Letter, we propose a system of  two-component BEC trapped
 in a harmonic potential \cite{cornell}, subject to a periodic
 coupling (successive kicks) between the two components. Our aim is
 two-fold: (1) To investigate the
 instability of the  BEC  system with a small perturbation on its system parameters;
 (2) To propose a possible experiment to directly detect the fidelity decay.

 The system we propose is a two-component spinor BEC
 confined in a harmonic trap with two internal states coupled by a
 near resonant pulsed radiation field\cite{cornell}. Within the
 standard rotating-wave approximation, the Hamiltonian can be cast
 into the form \cite{leggett},
 \begin{equation}
 \hat{H}=\mu (\hat{a}_1^{\dagger }\hat{a}_1-\hat{a}_2^{\dagger }\hat{a}%
 _2)+g(\hat{a}_1^{\dagger }\hat{a}_1-\hat{a}_2^{\dagger }\hat{a}_2)^2+K\delta
 _T(t)(\hat{a}_1^{\dagger }\hat{a}_2+\hat{a}_2^{\dagger }\hat{a}_1)  \label{1}
 \end{equation}
 where $K$ is the coupling strength between the two internal states, $g$ is the
 interaction strength, and $\mu$ is the difference between  the chemical potentials of
 two components. $\hat{a}_1,\hat{a}_1^{\dagger },\hat{a}_2$ and $\hat{a}%
 _2^{\dagger }$ are boson annihilation and creation operators for the two
 components, respectively. $\delta _T(t)=\sum_n\delta (t-nT)$ means that
 the radiation field is only turned on at certain discrete moments, i.e., integral multiples  of
 the period $T$.
 Writing the above
 Hamiltonian in terms of the angular momentum operators \cite{anglin}, $\hat{L%
 }_x=\frac{\hat{a}_1^{\dagger }\hat{a}_2+\hat{a}_2^{\dagger }\hat{a}_1}2,$ $%
 \hat{L}_y=\frac{\hat{a}_1^{\dagger }\hat{a}_2-\hat{a}_2^{\dagger }\hat{a}_1}{%
 2i},$ $\hat{L}_z=\frac{\hat{a}_1^{\dagger }\hat{a}_1-\hat{a}_2^{\dagger }%
 \hat{a}_2}2,$ we have $\hat{H}=\mu \hat{L}_z+g\hat{L}_z^2+K\delta _T(t)\hat{L%
 }_x$.  The Floquet operator depicting  the quantum evolution in one period takes the following form,
 \begin{equation}
 \hat{U}=\exp [-i(\mu \hat{L}_z+g\hat{L}_z^2)T]\exp (-iK\hat{L}_x).  \label{2}
 \end{equation}
 The Hilbert space is spanned by the eigenstates of $\hat{L}_z$, $|l\rangle $%
 , with $l=-L,-L+1,\ldots ,L$, where $L=N/2$ and $N$ is the total number of
 atoms. In the above expression and henceforth, the Planck constant is set to
 unit.

\begin{figure}[t]
\includegraphics[width=\columnwidth]{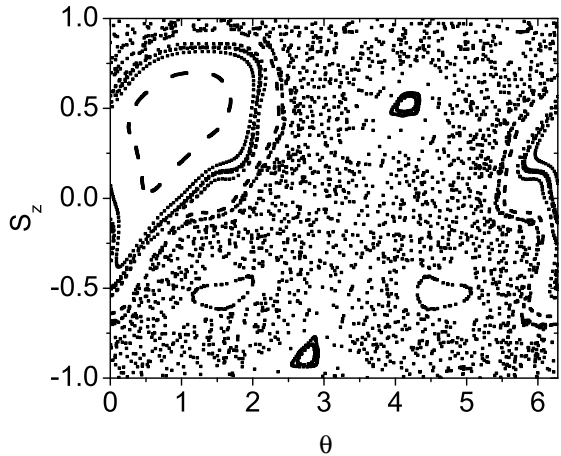} \vspace{-1.0cm}
\caption{ Stroboscopic plots of the orbits for $g_c=1,K=2$ where x-axis $\theta$ is the azimuthal angle. It shows
one big island and four small islands. Inside the islands motions are stable, outside the islands motions are
mainly unstable or chaotic. Here and in the following figures, $\mu =T=1$. } \label{fig-phase}
\end{figure}

 The above system has a classical counterpart in the limit $N\to \infty $,
 describing a spin on a Bloch sphere with $S_i=\frac 1L<\hat{L}%
 _i>,(i=x,y,z)$. The classical Hamiltonian takes the form, $H=\mu
 S_z+g_cS_z^2+K\delta _T(t)S_x$, where $g_c=gL$. The equations $\dot{S}%
 _i=[S_{i,}H]_{cl},(i=x,y,z)$ determine the motion of the centers of coherent quantum
 wavepackets and  the
 quantum fluctuation is  ignored, (i.e., equivalent to the mean-field Gross-Pitaeviskii equation
 without considering a total phase\cite{liukick}). They can be solved analytically:
 the free evolution between two consecutive kicks corresponds
 to a rotation around $S_z$ axis with the angle $(\mu +2g_cS_z)T$,
 and the periodic kicks added at times $nT$ give
 rotation around the $S_x$ axis with the angle $K$.

 Dynamic motion of the classical system is classified by the
 magnification of its initial deviation. An exponential increase in
 time of the deviation means dynamical instability or chaotic
 motion\cite{dyinst}, causing rapid
proliferation of thermal particles\cite{liukick}.
 Quantitatively, one can calculate the (maximum)
 Lyapunov exponent, $\lambda = \lim_{t \to \infty } \frac{1}{t}\ln \frac{| \delta x (t)|%
 }{|\delta x(0) |}$, with $|\delta x(t)|$ denoting distance in
 phase space. The exponent is positive for unstable motion, and
 tends to zero if the orbit is stable. Usually, phase
 space is a mixture of chaotic orbits and quasi-periodic (stable)
 orbits, as is shown in Fig.1, where it is clearly seen one big island
 and four small islands; inside the islands motions are
 periodic or quasi-periodic, outside the islands motions are
 mainly chaotic.

\begin{figure}[b]
\includegraphics[width=\columnwidth]{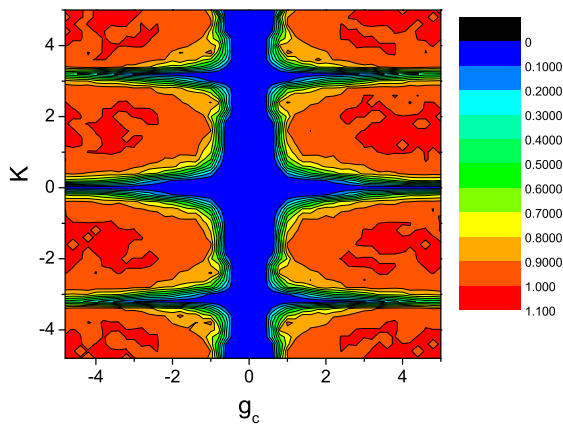} \vspace{-0.8cm}
\caption{ Contour plot of the fraction of chaotic orbits in phase
space,  with respect to  system parameters. } \label{fig-spin}
\end{figure}

\begin{figure}[b]
\includegraphics[width=\columnwidth]{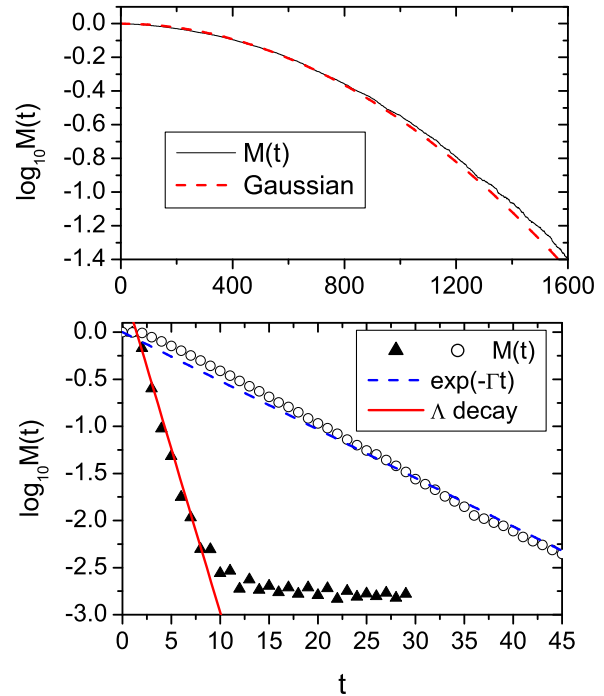} \vspace{-1.2cm}
\caption{ Fidelity decay in a fully chaotic case of $K=2, g_c=4$. Upper panel: An example of Gaussian decay at
small perturbation, with $ L=100, \epsilon =2 \times 10^{-4}$, obtained from one initial coherent state. The
dashed curve is the Gaussian function of form $\exp \left ( -1.3\times 10^{-6} t^2 \right )$. Lower panel:
Circles: fidelity decay with an  intermediate perturbation (above $\epsilon_p$), where the average has been made
over 20 initial coherent states chosen randomly. The dashed line is the exponential decay with $\Gamma = 0.12$.
Triangles: Strong perturbation regime where the fidelity decays as $e^{-\Lambda t}$ for a short time ($t<10$) and
then saturates, with $L=500, \epsilon =1 \times 10^{-2}$, and average over 1000 initial coherent states.
 Here $\Lambda=0.8$ is independent of perturbation
 strength\cite{wclp04}.}
\label{fig-chaotic-all}
\end{figure}

 The total relative area occupied by chaotic orbits, (as clearly seen in Fig.1),
  depends on system parameters ($g_c,K$), and
  can be used to characterize the degree of mixture. It has
 been
 obtained by calculating the Lyapunov exponent of orbits with
 initial points randomly scattered in the whole phase space. The result is   shown in Fig.2.
 We see that the
 integrable cases mainly concentrate on  the vertical line where the interaction  strength
 vanishes,
 and  on  the horizontal lines where the coupling strength is a multiple  of  $\pi
 $. This fact
 indicates that both nonlinearity term and the kick strength are essential in inducing
 chaos. The deep red areas in Fig.2  give the parameter regime for the
 system where the  phase space
 is  full of
 unstable (chaotic) orbits.

 Now we turn to the quantum system and trace the fidelity (Loschmidt echo) $M(t)$, defined as
 \begin{equation}
 M(t=nT)=\langle \Phi _0|
 \left ( \hat{U}_\epsilon ^{\dagger } \right )^n \circ
 \left ( \hat{U} \right )^n
 |\Phi _0\rangle ,
 \label{3} \end{equation}
 where the initial state $|\Phi _0\rangle $ is chosen as a coherent state, $%
 |\Phi _0\rangle =e^{\alpha ^{*}L_{+}-\alpha L_{-}}|-L\rangle $, with $\alpha
 =\frac{\pi -\theta }2e^{-\mathrm{i}\varphi }$. A small perturbation on the
 Hamiltonian is added by changing $K\to K+\varepsilon $, with $\hat{U}\to
 \hat{U}_\epsilon $. In this system, the effective Planck constant $\hbar _{%
 \mathrm{eff}}=1/L$.

\begin{figure}[t]
\includegraphics[width=\columnwidth]{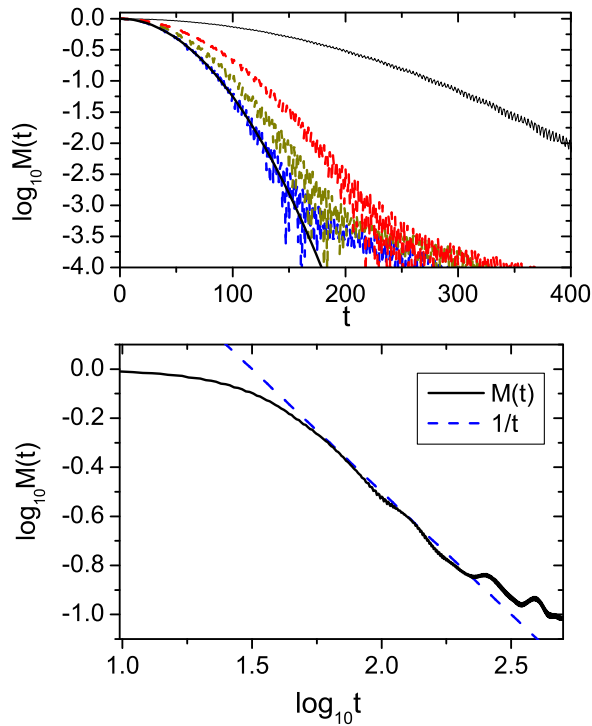} \vspace{-1.cm}
\caption{ Fidelity decay in a classically nearly integrable case,
with $K=2, g_c=0.2, L=100$, and $\epsilon = 0.003$. Upper panel:
Fidelity of four randomly chosen initial coherent states, with the
smooth solid curve being the Gaussian fit to one of them. Lower
panel: Averaged fidelity, with average performed over 50 initial
coherent states. } \label{fig-g02-all}
\end{figure}

 We discuss fidelity decay in three typical situations, in
 which the corresponding classical system is fully chaotic,
 near-integrable, and mixed, respectively. The corresponding
 parameters are picked up from Fig.2.

 For the parameters $K=2, g_c=4$, from Fig.2 we know that the phase
 space is fully chaotic. Because of the ergodicity of the chaotic orbits,
 fidelity decay is expected to be independent on
 the initial condition. However, it strongly depends on the perturbation strength.
 For a small perturbation, fidelity shows a
 slow Gaussian decay (upper panel in Fig.3).
 With increasing perturbation strength,
 one meets a border $\epsilon_p \sim 1/L^{3/2}$,
 at which the typical transition matrix
 element of perturbation between quasi-energy eigenstates becomes larger than
 the average level spacing.
 With the intermediate perturbation above the border (lower panel in Fig.3),
 the fidelity decays in an exponential
 way, where the  decay rate $\Gamma$ is the  function of  the interaction strength
 and
  the classical action diffusion constant\cite{CT02}. With strong
perturbation, the fidelity decays faster
 and finally saturates at some perturbation-independent decay rate \cite{wclp04} (lower panel in Fig.3).

 From the above discussions and calculations we see,
 in practical applications of  the BEC,
 the perturbation border $\epsilon_p $ gives a up-limit for the
 perturbation strength that is  tolerable, in order to avoid low fidelity.

 As we choose parameters as $K=2, g_c=0.2$, the classical system is
 nearly integrable where the phase space is full of periodic and
 quasi-periodic orbits. We found Gaussian decay for
 the  fidelity of single initial coherent states, with a strong dependence of decaying rate on the choice of initial
 condition \cite{PZ02}. However, after averaging over the whole phase space,
  we found that  the fidelity decay can be well fitted by a inverse power law  $1/t$
  (see Fig.~\ref{fig-g02-all}).
 In this case,
 for the quantum evolution of  initial coherent states, high fidelity can be
 expected because the fidelity has a power law decay on average.

 \begin{figure}[!t]
 \includegraphics[width=\columnwidth]{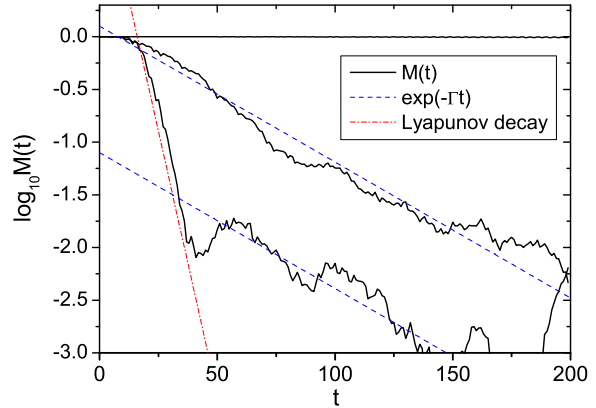} \vspace{-1.0cm}
 \caption{ Fidelity decay in the mixed system whose classical phase space
 structure is shown in Fig.~\ref{fig-phase}. $L=500$ and $\epsilon = 6\times 10^{-4}$, the intermediate perturbation.
 The non-decaying solid line is the fidelity of an initial coherent state
 lying within the largest regular region. The other two solid curves
 correspond to fidelity of two initial coherent states lying in the chaotic
 region of the classical system.  One of them has an exponential decay
 with $\Gamma =0.03$, as expected.
 Unexpectedly, the other one first has
 a fast Lyapunov decay $e^{-\lambda t}$, with $\lambda $ being the Lyapunov
 exponent, then follows the  exponential decay as the first one. }
 \label{fig-fid-g1}
 \end{figure}

 Now we turn to the mixed case, which is more complicated than the previous two cases.
 It is usually expected that fidelity decay of initial
 coherent states lying in regular regions would be similar to that in a
 nearly integrable system, and that from chaotic regions be similar to that
 in a chaotic system. However, we found that this naive picture is not exact. As shown in Fig.~\ref{fig-fid-g1},
 for initial states  from  both irregular and regular regions,
 the behavior of fidelity may be  quite different from those in the fully
 chaotic case and in the nearly-integrable case as shown in Figs. 3 and 4. We
 concentrate our discussions on the case in which  initial coherent
 states lie within the largest regular island. We found that their fidelity almost
 has  no decay up to time $t=200$, quite different from the initial-condition-dependent
 Gaussian decay shown in Fig.~\ref{fig-g02-all}
 for a nearly integrable system. Note that the quantum
 perturbation strength is chosen to be in the intermediate perturbation regime
 in Figs. \ref{fig-g02-all} and \ref{fig-fid-g1}. This phenomenon of high fidelity cannot be
 explained by means of expanding the coherent states in the eigenstates $|\alpha \ra $ of
 the system \cite{Peres84}, since the values of the participation function
 of the coherent states, defined by $1/(\sum_{\alpha } |\langle \alpha |\Phi_0 \rangle |^4 )$, is about 22.
 The principle of this  way of sustaining high fidelity in quantum
 evolution of an coherent state may be
  useful in applying  BEC in information processing.

\begin{figure}[!t]
\includegraphics[width=\columnwidth]{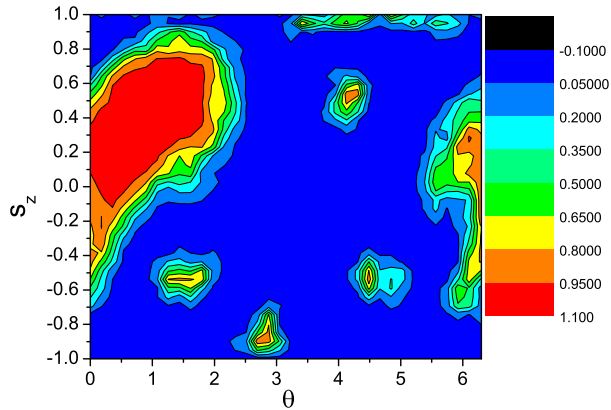} \vspace{-1.2cm}
\caption{ Contour plot of the fidelity $M(t)$ at $t=200$, for $%
K=2,g_c=1,L=500$, and $\epsilon =0.0006$. The initial quantum
states are cohenet states with corresponding ($S_z,\theta$). }
\label{fig-fid-phase-t200}
\end{figure}

 In order to have a knowledge of the global situation of fidelity decay in a
 mixed system, in Fig.\ref{fig-fid-phase-t200} we show a contour plot for $%
 M(t=200)$, with respect to initial  coherent states. With this figure at hand, in applying BEC to quantum
 information processing we may carefully choose the parameters to
 avoid the regimes of low fidelity. Moreover,  in surprise  we
 find  the  structure of fidelity plotting in Fig.6 quite similar
 to that of classical phase space in Fig.\ref{fig-phase}. This
 similarity
indicates a kind of connection between the dynamical instability
of the classical meanfield equation and the fidelity  of quantum
boson system, i.e., the dynamical instability regime of the
classical system usually corresponds to the low fidelity regime of
the quantum system. Moreover, inside  the islands (the large or
small) where the classical motions are dynamical stable with zero
Lyapunov exponent, the fidelity shows different behavior: The
fidelity in the large island of a mixed-type phase space is higher
than that in the small islands or  even that in near-integrable
case. This fact indicates that  the fidelity contains more
information about the system under a perturbation and therefore is
a more general quantity to describe the stability of the BEC.

 Experimentally, one can prepare two nearly identical two-component
 BECs by applying a strong blue detuning laser beam at the center
 of the two-component BEC with prepared initial state $\psi \chi $,
 where $\psi $ is the external state and $\chi $ is the internal
 state \cite{ketterle}. The strength of laser beam is strong enough,
 so that the tunnelling between the two condensates is negligible. Then,
 two pulsed radiation fields with slight different strengths are
 applied to the two condensates and kick the internal state to $\chi
 _i$ ($i=1,2$) without important change of the external states
 $\psi _i$ \cite{cornell}. After certain numbers of kicks, the
 radiation field and the strong blue detuning laser beam are turned
 off simultaneously and two BECs begin to interfere. The visibility
 of the interference is governed by
 \begin{equation}
 I\varpropto \left| \psi _1\chi _1\right| ^2+\left| \psi _2\chi _2\right|
  ^2+2Re\left( \psi _1^{*}\psi _2\chi _1^{*}\chi _2\right) .  \label{4}
 \end{equation}
 Clearly, high fidelity of the two internal states corresponds to high
 visibility of the interference.

The work was supported in part by a Faculty Research Grant of
National University of Singapore, the Temasek Young Investigator
Award of DSTA Singapore under Project Agreement POD0410553 (BL),
the Natural Science Foundation of China No.10275011 (WGW),
NSFC10474008, and the Natural Science Foundation of US.


\begin{thebibliography}{99}
\bibitem{lukin} M. D. Lukin, Rev. Mod. Phys. \textbf{75}, 457 (2003).
\bibitem{bec}  M.H.Anderson, et al, Science \textbf{269}, 198 (1995); K.Davis, et al,
Phys. Rev. Lett. \textbf{75}, 3969 (1995); C.C.Bradley, et al,
Phys. Rev. Lett. \textbf{75}, 1687 (1995).

\bibitem{dyinst}  S. Sinha and Y. Castin,Phys. Rev. Lett. \textbf{87}, 190402
(2001);P. Buonsante, R. Franzosi, and V. Penna,Phys. Rev. Lett.
\textbf{90}, 050404 (2003); C. P. Search, H. Pu, W. Zhang, and P.
Meystre,Phys. Rev. A \textbf{65}, 063615 (2002).

\bibitem{Landinst}  L. J. Garay, J. R. Anglin, J. I. Cirac, and P.
Zoller,Phys. Rev. Lett. \textbf{85}, 4643-4647 (2000); B. Wu and
Q. Niu, Phys. Rev. A \textbf{64}, 061603 (2001).

\bibitem{modins}  V. V. Konotop and M. Salerno,Phys. Rev. A \textbf{65}, 021602
(2002);G. Theocharis, Z. Rapti, P. G. Kevrekidis, D. J.
Frantzeskakis, and V. V. Konotop,Phys. Rev. A \textbf{67}, 063610
(2003); L. Salasnich, A. Parola, and L. Reatto,Phys. Rev. Lett.
\textbf{91}, 080405 (2003);L. D. Carr and J. Brand,Phys. Rev.
Lett. \textbf{92}, 040401 (2004).

\bibitem{smerzianglin}  J. R. Anglin and A. Vardi,Phys. Rev. A \textbf{64}, 013605
(2001);V. A. Yurovsky,Phys. Rev. A \textbf{65}, 033605 (2002);G.
P. Berman, A. Smerzi, and A. R. Bishop,Phys. Rev. Lett.
\textbf{88}, 120402 (2002);J. R. Anglin,Phys. Rev. A \textbf{67},
051601 (2003).

\bibitem{liukick}  Jie Liu, Biao Wu, and Qian Niu
Phys. Rev. Lett. \textbf{90}, 170404 (2003); C. Zhang, J. Liu, M.
G. Raizen, and Q. Niu,Phys. Rev. Lett. \textbf{92}, 054101 (2004)
.

\bibitem{quancomp}  M.A.~Nielsen and I.L.~Chuang, \emph{Quantum Computation
and Quantum Information} (Cambridge University Press, Cambridge,
2000).
\bibitem{JP01}
G.~Benenti and G.~Casati, Phys. Rev. E \textbf{65}, 066205(2002);
W.Wang and B.Li, Phys.~Rev.~E
\textbf{66}, 056208 (2002);W.Wang, G.Casati, and B.Li, Phys.~Rev.~E \textbf{%
69}, 025201(R)(2004).


\bibitem{CT02} N.~R.~Cerruti and
S.~Tomsovic, Phys.~Rev.~Lett. \textbf{88}, 054103 (2002).

\bibitem{JAB02} Chuanwei Zhang, Jie Liu, Mark G. Raizen, and Qian Niu
Phys. Rev. Lett. \textbf{93}, 074101 (2004).

\bibitem{PZ02}  T.Prosen and M.\v{Z}nidari\v{c}, J.Phys.A \textbf{35}, 1455
(2002).

\bibitem{wclp04}  Wen-ge Wang, G.~Casati, Baowen Li, and
T.~Prosen, Phys. Rev. E \textbf{71}, 186503 (2005).

\bibitem{cornell}  M. R. Matthews, B. P. Anderson, P. C. Haljan,
D. S. Hall, M. J. Holland, J. E. Williams, C. E. Wieman, and E. A.
Cornell,  Phys. Rev. Lett. \textbf{83}, 3358 (1999).

\bibitem{leggett}  A.J.Leggett, Rev. Mod. Phys. \textbf{73}, 307 (2001).

\bibitem{anglin}  A. Vardi and J. R. Anglin, Phys. Rev. Lett. \textbf{86}, 568-571
(2001).

\bibitem{Peres84}  A.~Peres, Phys.~Rev.~A \textbf{30}, 1610 (1984).

\bibitem{ketterle}  Y. Shin, M. Saba, T. A. Pasquini, W. Ketterle, D. E. Pritchard, and A. E. Leanhardt , Phys. Rev. Lett. \textbf{92}, 050405
(2004).

\end{thebibliography}
\end{document}